%% file: ftscs2012-wang-davies.tex
\documentclass[submission,copyright,creativecommons]{eptcs}
\providecommand{\event}{FTSCS 2012}
                   
\usepackage{graphicx}
\usepackage{url} \usepackage{hyperref} \usepackage{multirow}
\usepackage{graphicx} \usepackage{hyperref}
\usepackage[zed]{csp} \input{macros}  
	
\usepackage{titlesec}

\renewenvironment{thebibliography}[1]{%
  \begin{oldthebibliography}{#1}%
    \raggedright
    \setlength{\parskip}{.3ex}%
    \setlength{\itemsep}{.3ex}%
  }%
  {%
  \end{oldthebibliography}%
  }

\titlespacing*{\subsection}{0pt}{*0.5}{*0.5}
\titlespacing*{\subsubsection}{0pt}{*0.5}{*0.5}    

\DefineVerbatimEnvironment{haskell}{Verbatim}{frame=none, framesep=10mm, fontsize=\footnotesize, commandchars=\\\{\}}

\def\bfword#1{\texttt{\textbf{#1}}}

\title{Formal Model-Driven Engineering: \\ Generating Data and Behavioural Components}   
                           
\author{Chen-Wei Wang
\institute{McMaster Centre for Software Certification, \\ McMaster University \\ Hamilton, Canada L8S 4K1}
\email{jackie@cse.yorku.ca}
\and
Jim Davies
\institute{Department of Computer Science, \\ University of Oxford \\ Oxford, United Kingdom OX1 3QD}
\email{jim.davies@cs.ox.ac.uk}
}                            

\def\titlerunning{Formal Model-Driven Engineering}
\def\authorrunning{C.-W. Wang and J. Davies}                         

\begin{document}                        
	
\maketitle

\begin{abstract} 
  Model-driven engineering is the automatic production of software
  artefacts from abstract models of structure and functionality.  By
  targeting a specific class of system, it is possible to automate
  aspects of the development process, using model transformations and
  code generators that encode domain knowledge and implementation
  strategies.  Using this approach, questions of correctness for a
  complex, software system may be answered through analysis of
  abstract models of lower complexity, under the assumption that the
  transformations and generators employed are themselves correct.
  This paper shows how formal techniques can be used to establish the
  correctness of model transformations used in the generation of
  software components from precise object models.  The source language
  is based upon existing, formal techniques; the target language is
  the widely-used SQL notation for database programming.  Correctness
  is established by giving comparable, relational semantics to both
  languages, and checking that the transformations are
  semantics-preserving.
\end{abstract}

\bigskip

\section{Introduction}

Our society is increasingly dependent upon the behaviour of complex
software systems.  Errors in the design and implementation of these
systems can have significant consequences.  In August 2012, a `fairly
major bug' in the trading software used by Knight Capital Group lost
that firm \$461m in 45 minutes~\cite{knight}.  A software glitch in
the anti-lock braking system caused Toyota to recall more than 400,000
vehicles in 2010~\cite{toyota}; the total cost to the company of this
and other software-related recalls in the same period is estimated at
\$3bn.  In October 2008, 103 people were injured, 12 of them
seriously, when a Qantas airliner~\cite{qantas} dived repeatedly as
the fly-by-wire software responded inappropriately to data from
inertial reference sensors.  A modern car contains the product of over
100 million lines of source code~\cite{Charette2009}, and in the
aerospace industry, it has been claimed that ``the current development
process is reaching the limit of affordability of building safe
aircraft''~\cite{Feiler2010}.

The solution to the problems of increasing software complexity lies in
the automatic generation of correct, lower-level software from
higher-level descriptions: precise \emph{models} of structure and
functionality.  The intention is that the same generation process
should apply across a class of systems, or at least multiple versions
of the same system.  Once this process has been correctly implemented,
we can be sure that the behaviour of the generated system will
correspond to the descriptions given in the models.  These models are
strictly more abstract than the generated system, easier to understand
and update, and more amenable to automatic analysis.  This
\emph{model-driven} approach~\cite{Frankel2003} makes it easier to
achieve correct designs and correct implementations.  Despite the
obvious appeal of the approach, and that of related approaches such as
domain-specific languages~\cite{Deursen2000} and software product
lines~\cite{Pohl2005}, much of the code that could be generated
automatically is still written by hand; even where precise, abstract
specifications exist, their implementation remains a time-consuming,
error-prone, manual process.

The reason for the delay in uptake is simple: in any particular
development project, the cost of producing a new language and matching
generator, is likely to exceed that of producing the code by hand.  As
suitable languages and generators become available, this situation is
changing, with significant implications for the development of
complex, critical, software systems.  In the past, developers would
work to check the correctness of code written in a general-purpose
programming language, such as C or Ada, against natural language
descriptions of intended functionality, illuminated with diagrams and
perhaps a precise, mathematical account of certain properties.  In the
future, they will check the correctness of more abstract models of
structure and behaviour, written in a range of different,
domain-specific languages; and rather than relying upon the
correctness of a single, widely-used compiler, they will need to rely
upon the correctness of many different code generators.  The
correctness of these generators, usually implemented as a sequence of
model transformations, is thus a major, future concern.

In this paper, we present an approach to model-driven development that
is based upon formal, mathematical languages and techniques.  The
objective is the correct design and implementation of components with
complex state, perhaps comprising a large number of inter-related data
objects.  The approach is particularly applicable to the iterative
design and deployment of systems in which data integrity is a primary
concern.  The modelling language employed has the familiar, structural
features of object notations such as UML---classes, attributes, and
associations---but uses logical predicates to characterise operations.
An initial stage of transformation replaces these predicates with
guarded commands that are guaranteed to satisfy the specified
constraints: see, for example,~\cite{Welch2008}.  The focus here is
upon the subsequent generation of executable code, and the means by
which we may prove that this generation process is correct.

The underlying thesis of the approach is that the increasing
sophistication of software systems is often reflected more in the
complexity of data models than in the algorithmic complexity of the
operations themselves.  The intended effect of a given event or action
is often entirely straightforward.  However, the intention may be only
part of the story: there may be combinations of inputs and
before-states where the operation, as described, would leave the
system in an inconsistent after-state; there may be other attributes
to be updated; there may be constraints upon the values of other
attributes that need to be taken into account.  Furthermore, even if
the after-state is perfectly consistent, the change in state may have
made some other operation, or sequence of operations, inapplicable.

Fortunately, where the intended effect of an operation upon the state
of a system is straightforward, it should be possible to express this
effect as a predicate relating before and after values \emph{and}
generate a candidate implementation.  Using formal techniques, we may
then calculate the domain of applicability of this operation, given
the representational and integrity constraints of the data model.  If
this is smaller than required, then a further iteration of design is
called for; if not, then the generated implementation is guaranteed to
work as intended.  In either case, code may be generated to throw an
exception, or otherwise block execution, should the operation be
called outside its domain.  Further, by comparing the possible
outcomes with the calculated domains of other operations, we can
determine whether or not one operation can affect the availability of
others.

The application of formal techniques at a modelling level---to
predicates, and to candidate implementations described as abstract
programs---has clear advantages.  The formal semantics of a modern
programming language, considered in the context of a particular
hardware or virtual machine platform, is rich enough to make
retrospective formal analysis impractical.  If we are able to
establish correctness at the modelling level, and rely upon the
correctness of our generation process, then we may achieve the level
of formal assurance envisaged in new standards for certification: in
particular, DO-178C~\cite{DO178C}.  We show here how the correctness
of the process can be established: in Section~\ref{framework}, we
present the underlying semantic framework; in
Section~\ref{sql:trans:path}, the translation of expressions; in
Section~\ref{sql:trans:assignment}, the implementation of operations;
in Section~\ref{sql:semantics}, the approach to verification.

\section{Preliminaries}~\label{background} \vspace*{-\baselineskip}

\noindent
The $\Booster$ language~\cite{Davies2005} is an object modelling
notation in which model constraints and operations are described as
first-order predicates upon attributes and input values.  Operations
may be composed using the logical combinators: conjunction,
disjunction, implication, and both flavours of quantification.  They
may be composed also using relational composition, as sequential
phases of a single operation or transaction.  The constraints
describing operations are translated automatically into programs
written in an extended version of the Generalised Substitution
Language (GSL), introduced as part of the B Method~\cite{Abrial1996}.
There may be circumstances under which a program would violate the
model constraints, representing business rules, critical requirements,
or semantic integrity properties.  Accordingly, a guard is calculated
for each operation, as the weakest precondition for the corresponding,
generated program to maintain the model constraints.  The result is an
abstract program whose correctness is guaranteed, in a language
defined by the following grammar:

\vspace{7pt}

{\small \noindent
$\begin{array}{lclllcl}
  Substitution &::=& \Pskip 
               & | & \ldata \Path \rdata \PASS \ldata Expression \rdata \\
               & | & \ldata Predicate \rdata \PGUARD \ldata Substitution \rdata 
               & | & \ldata Substitution \rdata \PAND \ldata Substitution \rdata \\
               & | & \ldata Substitution \rdata \PTHEN \ldata Substitution \rdata 
               & | & \ldata Substitution \rdata \POR \ldata Substitution \rdata \\
               & | & \PFORALL \ldata Variable \rdata : \ldata Expression \rdata @ \ldata Substitution \rdata \\ 
               & | & \PEXISTS \ldata Variable \rdata : \ldata Expression \rdata @ \ldata Substitution \rdata 
\end{array}$} 
                
\vspace{7pt}

Here, the usual notation of assignable variables is replaced with
\textit{paths}, each being a sequence of attribute names, using the
familiar object-oriented `dot' notation as a separator.  $Predicate$
and $Expression$ represent, respectively, first-order predicates and
relational and arithmetic expressions.  $\Pskip$ denotes termination,
$\PASS$ denotes assignment, and $\PGUARD$ denotes a program guard: to
be implemented as an assertion, a blocking condition, or as (the
complement of) an exception.  $\POR$ denotes alternation, and
$\PEXISTS$ denotes selection: the program should be executed for
exactly one of the possible values of the bound variable.  Similarly,
$\PAND$ denotes parallel composition, with $\PFORALL$ as its
generalised form: all of the program instances should be performed, in
parallel, as a single transaction.  $\PTHEN$ denotes relational or
sequential composition.  Inputs and outputs to operations need not be
explicitly declared; instead, they are indicated using the decorations
$?$ and $!$ at the end of the attribute name.

These abstract programs are interpreted as operations at a component
applications programming interface (API), with the data model of the
component given by a collection of class and association declarations
in the usual object-oriented style.  The integrity constraints and
business rules for the data model can be given as predicates in the
same notation, or using the object constraint language (OCL) of the
Unified Modelling Language (UML)~\cite{Frankel2003}.    

\begin{figure*}[t] \centering 
  \includegraphics[width=\textwidth]{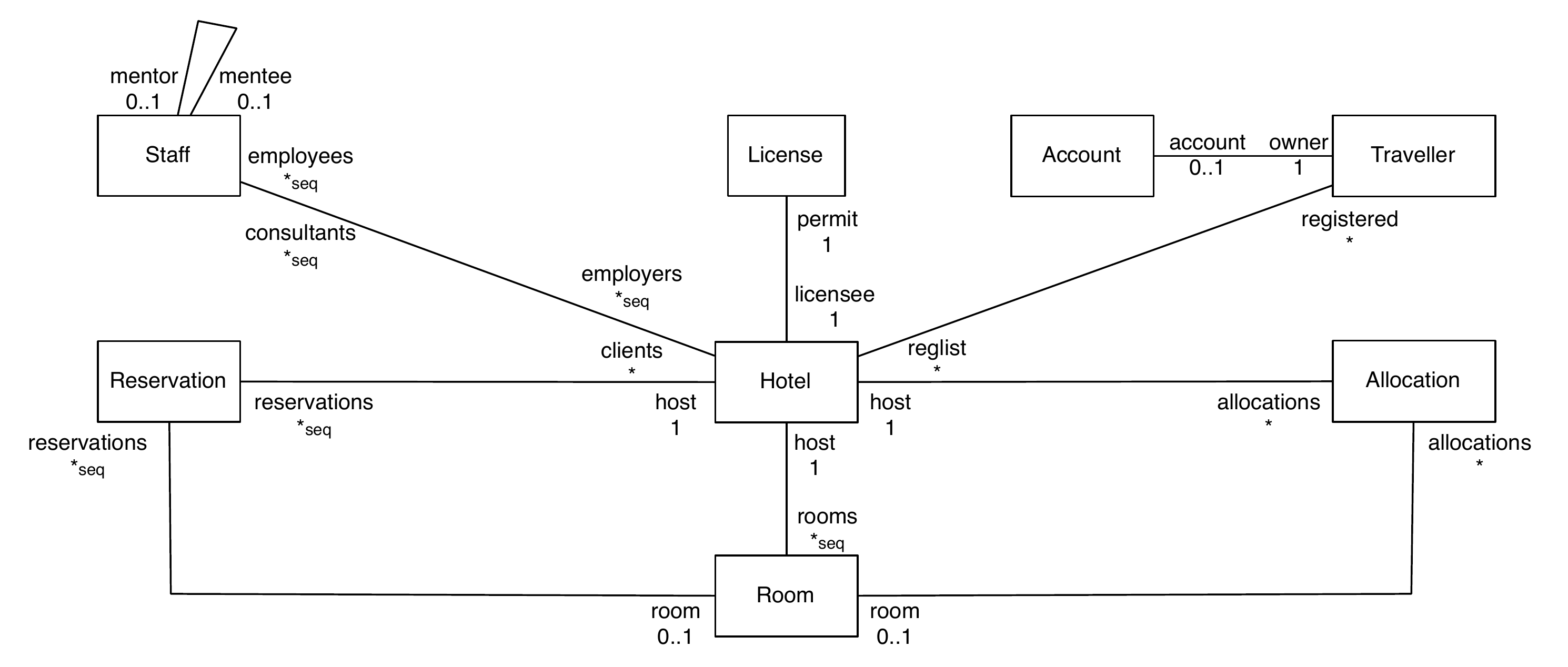} \vspace{-1cm}
  \caption{\small Hotel Reservation System (HRS)---Graph of Class
Associations}\label{context:fig:hrs}
\end{figure*}

As a simple, running example, consider the following description of
(a fragment of) the data model for a hotel reservations system 
\def\star{[*]}
\begin{Verbatim}[frame=none, framesep=10mm, fontsize=\footnotesize, commandchars=~\[\]]
  ~textbf[class] Hotel {                                    ~textbf[class] Reservation {
    ~textbf[attributes]                                       ~textbf[attributes]
      reservations : ~textbf[seq](Reservation.host) ~star }       host : Hotel.reservations }
\end{Verbatim}
A single hotel may be the \texttt{\textbf{host}} for any number of
reservations.  It may also be the \texttt{\textbf{host}} of a number
of rooms and allocations: see the class association
graph~\cite{Eriksson2003} of Figure~\ref{context:fig:hrs}.  The action
of creating a new reservation may be specified using a simple
operation predicate in the context of the \texttt{Hotel} class:
\begin{Verbatim}[frame=none, framesep=10mm, fontsize=\footnotesize, commandchars=~\[\]] 
  reserve { # allocations < limit & reservations' = reservations ^ <r!> & r!.room = m? }
\end{Verbatim}
This requires that a new reservation be created and appended to the
existing list, modelled as an ordered association from \texttt{Hotel}
to \texttt{Room}, and that the room involved is given by input
\texttt{m?}.  The operation should not be allowed if the number of
reservations in the system has already reached a specified
\texttt{limit}.

If the constructor operation predicate on \texttt{Reservation}
mentions a set of dates \texttt{dates?}, then this will be added as a
further input parameter.  We might expect to find also a constraint
insisting that any two different reservations associated with the same
room should have disjoint sets of dates, and perhaps constraints upon
the number of reservations that can held by a particular traveller for
the same date.  For the purposes of this paper, however, we will focus
simply upon the required, consequential actions and the description of
the operation as an abstract program. 

\begin{Verbatim}[frame=none, framesep=10mm, fontsize=\footnotesize, commandchars=~\[\]] 
  reserve {
       r! : extent(Reservation) & dates? : set(Date) & m? : extent(Room)
         & card(allocations) < limit
    ==>
       r!.dates := r!.dates \/ dates? || r!.status := "unconfirmed"      
    || r!.host := this || reservations := ins(reservations, #reservations + 1, r!)        
    || r!.room := m?   || m?.reservations := ins(m?.reservations, #m?.reservations + 1, r!)}
\end{Verbatim}                                                  

In this abstract program, the two reservations attributes, in the
hotel and room objects, are updated with a reference to the new
reservation, the dates attribute of the new reservation is updated to
include the supplied dates, and the status attribute is set to
\verb|"unconfirmed"|, presumably as a consequence of the constructor
predicate for the \texttt{Reservation} class. 

\section{A Unified Implementation and Semantic
  Framework}~\label{framework}

\vspace*{-\baselineskip}

\begin{figure*}[htp]
  \centering %
  \includegraphics[width=.84\textwidth]{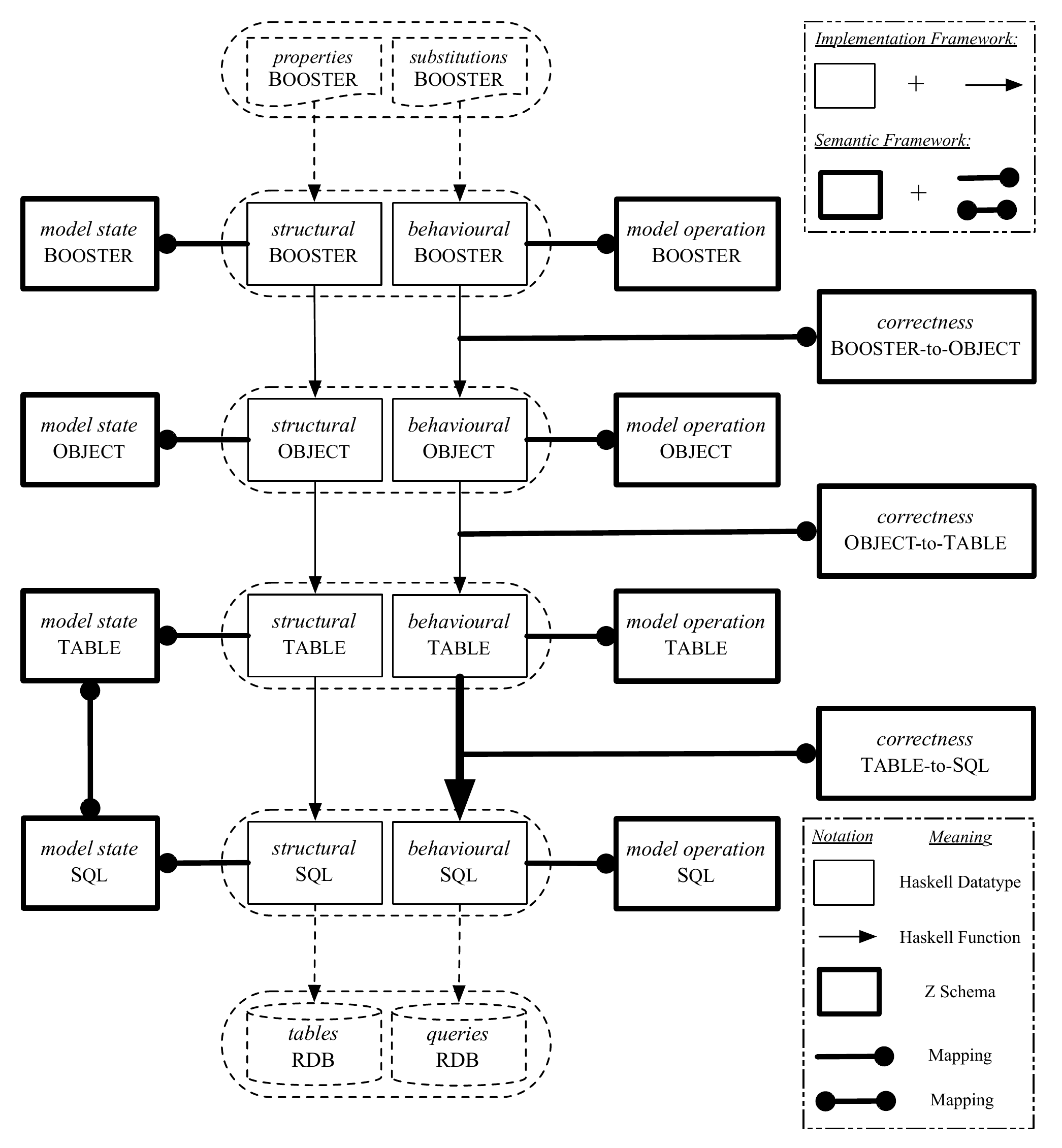} 
  \caption{\small $\Booster$ Model to $\Sql$ Database: Implementation \& Semantic Framework}
  \label{fig:sql:pipeline}  
\end{figure*}

\noindent
To illustrate our formal, model-driven approach, we will consider the
case in which the target is a relational database platform.  The above
program would then be translated into a SQL query, acting on a
relational equivalent of our original object model.  The
transformations can be described using the Haskell~\cite{Bird1998}
functional programming language: in the diagram of
Figure~\ref{fig:sql:pipeline}, thin-lined, unshaded boxes represent to
denote Haskell program data types, and thin arrows the executable
transformations between them.  These constitute an
\textit{implementation} framework.  The thick-lined, shaded boxes
denote the relational semantics of corresponding data types, thick
lines with circles at one end the process of assigning a formal
meaning, and arrows with circles at each end the relationship between
formalised concepts.  These constitute a corresponding
\textit{semantic} framework for establishing correctness.

Four kinds of models are involved in our transformation pipeline: 1) a
\underline{$\Booster$ model}, in extended GSL notation, generated from
the original predicates; 2) an \underline{$\Object$ model}
representing an object-oriented relational semantics for that model;
3) an intermediate \underline{$\Table$ model} reflecting our
implementation strategy; 4) a \underline{$\Sql$ model} expressed in
terms of tables, queries, and key constraints.  A final model-to-text
transformation will be applied to generate a well-formed $\Sql$
database schema.  

We use Haskell to define metamodels of model structures and operations
as data types.  Our transformations are then defined as Haskell
functions: from $\Booster$ to $\Object$, then to $\Table$, and finally
to $\Sql$.  Our relational semantics is most easily described using
the Z notation~\cite{Woodcock1996}.  Other formal languages with a
transformational semantics would suffice for the characterisation of
model and operation constraints, but Z has the distinct advantage that
each operation, and each relation, may be described as a single
predicate: rather than, for example, a combination of separate pre-
and post-conditions; this facilitates operation composition, and hence
a compositional approach to verification. 

\section{Path \& Expression Transformation}~\label{sql:trans:path} 

\vspace*{-\baselineskip}
                                                                          
\begin{figure}[htp]
  \centering
  \resizebox{.9\textwidth}{!}{ 
  \includegraphics{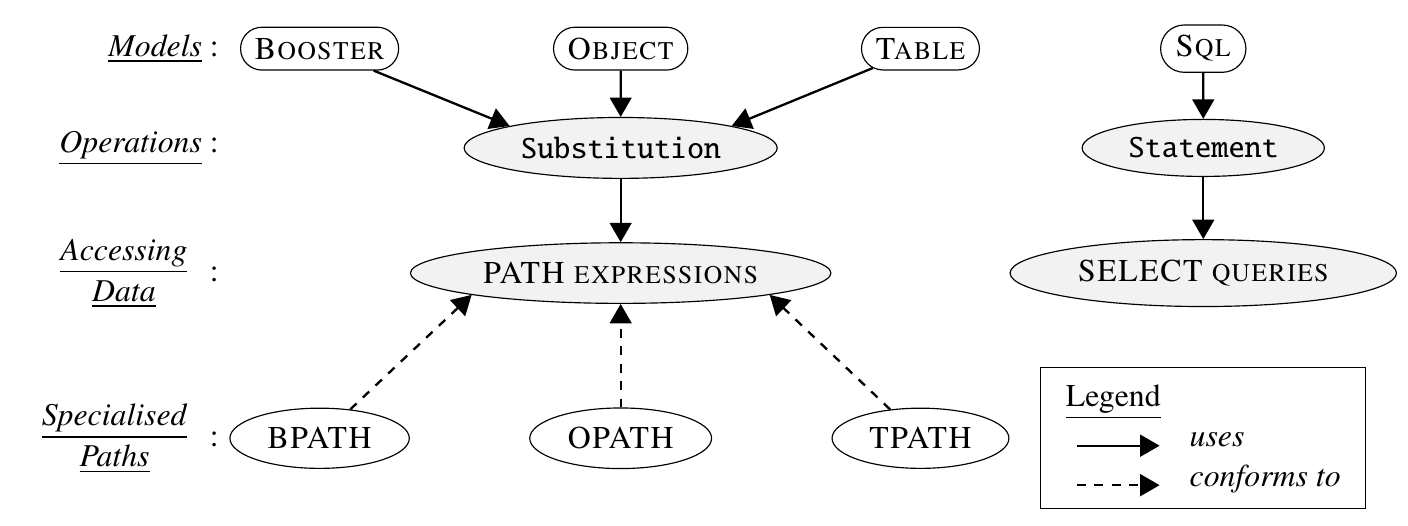}}
\vspace{-.5cm} 
  \caption{\small Datatypes of Behavioural Models}
  \label{fig:sql:imp-proof}    
\vspace{-.2cm}
\end{figure}

\noindent 
The descriptions of operations in the $\Booster$, $\Object$, and
$\Table$ models are all written in the GSL language; the difference
between them lies in the representation of attribute and association
references.  Instead of creating three versions of a language type
{\small \verb|Substitution|}, one for each of the reference notations,
we employ a type {\small \verb|PATH|} as a generic solution: see
Figure~\ref{fig:sql:imp-proof}.

\noindent 
We define 
\begin{haskell}
  data PATH = \textbf{BPath} BPATH | \textbf{OPath} OPATH | \textbf{TPath} TPATH
\end{haskell}  
where \bfword{BPath}, \bfword{OPath}, and \bfword{TPath} are type
constructors. A $\Booster$ model path (of type \word{BPATH}) is
represented as a sequence $\langle a_1, \dots, a_n \rangle$ of name
references to attributes/properties. We will refer to this range $1 \upto n$ of indices for explaining the corresponding $\Object$ and $\Table$ model paths.     

We consider structures of the types \word{OPATH} and \word{TPATH} in detail. Paths of type \word{OPATH} are used to indicate explicitly which properties/classes are accessed, along with its chain of navigation starting from the current class.
\begin{haskell}
  data OPATH         = \textbf{BaseOPath} REF_START        | \textbf{RecOPath} OPATH TARGET             
  data REF_START     = \textbf{ThisRef} BASE               | \textbf{SCRef}    IDEN_PROPERTY EXPRESSION BASE
  data TARGET        = \textbf{EntityTarget} IDEN_PROPERTY | \textbf{SCTarget} IDEN_PROPERTY EXPRESSION 
  data BASE          = \textbf{ClassBase} N_CLASS          | \textbf{SetBase} N_SET | \textbf{IntBase} | \textbf{StrBase}
  type IDEN_PROPERTY = (N_CLASS, N_ATTRIBUTE)
\end{haskell}
An object path is a left-heavy binary tree, where the left-most child refers to its starting reference and all right children represent target classes/properties that are accessed. The starting reference of an object path---which denotes access to, e.g~the current object, an element of a sequence-valued property through indexing, etc.---provides explicit information about the base type of that reference. All intermediate and the ending targets of an object path contextualise the properties with their enclosing classes (i.e.~{\small \verb|IDEN_PROPERTY|}).  

For each context path $\langle a_1,
\dots, a_i \rangle$, where ($1 \leq i \leq n - 1$), an $\Object$ model
path (of type \word{OPATH}) identifies a target class $C$; if the
source $\Booster$ path is valid, then attribute $a_{i + 1}$ must have
been declared in $C$.

\smallskip

\noindent\textit{Example object path.} As an example of how the transformation on paths works in practice, consider the {\small
  \verb|Account|} class (Figure~\ref{context:fig:hrs} shown on
page~\pageref{context:fig:hrs}).  The path \word{this.owner.reglist}
denotes a list of registered hotels and has its {\small
  \verb|OPATH|} counterpart:
\begin{haskell}
  \textbf{RecOPath} (\textbf{RecOPath} (\textbf{BaseOPath} (ThisRef (\textbf{ClassBase} Account)))
                     (\textbf{EntityTarget} (Account, owner)))
           (\textbf{EntityTarget} (Traveller reglist))
\end{haskell}
where \bfword{RecOPath} and \bfword{BaseOPath} are constructors for,
respectively, recursive and base $\Object$ paths.
\bfword{EntityTarget} and \bfword{ClassBase} construct type
information about the three context paths:~\word{(Account)} for
\word{this}, \word{(Account, owner)} for \word{this.owner}, and
\word{Traveller, reglist} for \word{this.owner.reglist}.

On the other hand, we use a path of type {\small \verb|TPATH|} to indicate, for each navigation to a property in the $\Object$ model, the corresponding access to a table which stores that property.
\begin{haskell}
  data TPATH    = \textbf{BaseTPath}                  REF_START  
                | \textbf{RecTPath}     TPATH         T_ACCESS     
  data T_ACCESS = \textbf{ClassTAccess} IDEN_PROPERTY
                | \textbf{AssocTAccess} IDEN_PROPERTY
                | \textbf{SetTAccess}   IDEN_PROPERTY
                | \textbf{SeqTAccess}   IDEN_PROPERTY             -- retrieve all indexed components
                | \textbf{SeqTCAccess}  IDEN_PROPERTY EXPRESSION  -- retrieve an  indexed component
\end{haskell}
A table path is left-heavy (as is an \word{OPATH}), where the left-most child refers to its starting reference and all right children represent target tables that are accessed. The starting reference of a table path provides exactly the same information as its {\small \verb|OPATH|} counterpart (i.e.~\word{REF\_START}). All intermediate and the ending targets of a table path denote accesses to a variety of tables, predicated upon our implementation strategy. When the target property is sequence-valued, we distinguish between the two cases where one of its indexed components is to be accessed ({\small \verb|SeqTCAccess|}) and where all indexed components are to be accessed ({\small \verb|SeqTAccess|}).

For each attribute $a_i$, where ($1 \leq i \leq n$), a $\Table$ model
path (of type \word{TPATH}) recursively records which sort of table
(e.g.~class tables, association tables, or set tables) it is stored,
based on the target class of its context path.

\smallskip

\noindent\textit{Example table path.} The above
$\Object$ path has its {\small \verb|TPATH|} counterpart:
\begin{haskell}
  \textbf{RecTPath} (\textbf{RecTPath} (\textbf{BaseTPath} (ThisRef (\textbf{ClassBase} Account)))
                     (\textbf{AssocTAccess} (Account, owner)))
           (\textbf{AssocTAccess} (Traveller reglist))
\end{haskell}
where \bfword{RecTPath} and \bfword{BaseTPath} construct,
respectively, recursive and base $\Table$ paths. Properties {\small
  \verb|owner|} and {\small \verb|reglist|} are accessed in the two
corresponding association tables.              

\smallskip
         
\noindent\textit{Path transformation.} We now specify the above \word{OPATH}-to-\word{TPATH} transformation in Haskell:

\begin{haskell}
  objToTabPath :: OBJECT_MODEL -> PATH -> PATH
  objToTabPath om (\textbf{OPath} opath) = \textbf{TPath} (objToTabPath' om opath)
\end{haskell}  

\noindent where the first line declares a function
\word{objToTabPath}, and the second line gives its definition:
matching an input object model as \word{om} and an input path as
\word{(OPath opath)}, whereas the RHS constructs a new \word{PATH} via
\word{TPath}.  The transformation of object paths is given by
\begin{haskell}
  objToTabPath' om (\textbf{RecOPath} op tar) =
    \textbf{case} tar \textbf{of}
      \textbf{EntityTarget} (c, p) | (c, p) `elem` biAssoc'    om c -> \textbf{RecTPath} tp (AssocTAccess (c, p))
                          | (c, p) `elem` classTables tm c -> \textbf{RecTPath} tp (ClassTAccess (c, p))
                          | (c, p) `elem` setTables   tm   -> \textbf{RecTPath} tp (SetTAccess   (c, p))
                          | (c, p) `elem` seqTables   tm   -> \textbf{RecTPath} tp (SeqTAccess   (c, p))
      \textbf{SCTarget} (c, p) oe  -> \textbf{let} te = objToTabExpr om oe   \textbf{in} \textbf{RecTPath} tp (SeqTCAccess (c, p) te)
    \textbf{where} tm = objToTab      om
          tp = objToTabPath' om op
\end{haskell}      
where each condition specified between \word{|} and \word{->} denotes
a special case of the matched entity target, consisting of class
\word{c} and property \word{p}.  For example, the condition \word{(c,
  p) `elem` classTables tm c} denotes properties that are stored in
the table for class \word{c}.  

Each recursive object path is structured as ({\small
  \verb|RecOPath op tar|}), where {\small \verb|op|} is its prefix
(i.e.~context) of type {\small \verb|OPATH|}, which we recursively
transform into a table path equivalent {\small \verb|tp|}; and {\small
  \verb|tar|} is its target property.  For each given {\small
  \verb|tar|}, table access is determined by checking membership
against various domains: a bidirectional association will be accessed
by means of an association table.  If the target property is sequence-valued (i.e.~the case of \word{SCTarget}), it cannot be accessed for its entirety, but only one of its members through indexing.  The function \word{objToTabExpr} transforms the index expression \word{oe} that contains paths of type \word{OPATH} to \word{te} that contains paths of type \word{TPATH}. The function \word{objToTab}
transforms an object model \word{om} to an equivalent table model
\word{tm}.  

$\Sql$ database statements express paths via (nested) $\SELECT$
queries. For example, the above $\Table$ path has its $\Sql$ statement
counterpart:
\begin{haskell}
  \textbf{SELECT} (\textbf{VAR}   `reglist`)
         (\textbf{TABLE} `Hotel_registered_Traveller_reglist`)
         (\textbf{VAR}   `oid` = (\textbf{SELECT} (\textbf{VAR}   `owner`)
                                (\textbf{TABLE} `Account_owner_Traveller_account`)
                                (\textbf{VAR}    oid = VAR this)))
\end{haskell}
where {\small \verb|oid|} is the default column (declared as the
primary key) for each table that implements an association.  We can
show~\cite{WangThesis2012} by structural induction that the
transformation from $BPATH$ to $OPATH$, from $OPATH$ to $TPATH$, and
from $TPATH$ to $\SELECT$ statements are correct.

\smallskip 

\noindent\textit{Expression transformation.} We transform both predicates and expressions on $\Table$ model into
$\Sql$ expressions:
\begin{haskell}
  toSqlExpr  :: TABLE_MODEL -> Predicate  -> SQL_EXPR
  toSqlExpr' :: TABLE_MODEL -> Expression -> SQL_EXPR
\end{haskell}
Some transformations are direct
\begin{haskell}
  toSqlExpr tm (\textbf{And} p q) = toSqlExpr tm p \textbf{`AND`} toSqlExpr tm q  
\end{haskell}
whereas others require an equivalent construction: 
\begin{haskell}
  toSqlExpr' tm (\textbf{Card} e) | isPathExpr e = \textbf{SELECT} [\textbf{COUNT} (VAR oid)] (toSqlExpr' tm e) \textbf{TRUE} 
\end{haskell}

\section{Assignment Transformation}~\label{sql:trans:assignment}    

\vspace{-\baselineskip}    

\noindent The most important aspect of the model transformation is the handling
of attribute assignments and collection updates.  There are 36
distinct cases to consider, given the different combinations of
attributes and bidirectional (opposite) associations.  We present a
single, representative case in Table~\ref{tab:pattern}, for an
association between an optional attribute (multiplicity 0..1) and a
sequence-valued attribute (ordered with multiplicity *) .

{\small \begin{table*}[htp] 
\footnotesize 
\centering
\begin{tabular}{|l|c|l|l|}
\hline
Bi-Assoc. Decl. & \# & GSL Substitution & $\Sql$ Queries \\
\hline \hline                        
\begin{block}
\begin{block}
\textrm{seq}\textit{-to-}\textrm{opt} \\ 
\underline{\decl{class A}} \\
\decl{  bs:\ seq(B.ao)} 
\end{block}
\quad
\begin{block} 
	\\   
\underline{\decl{class B}} \\
\decl{  ao:\ [A.bs]}
\end{block}
\end{block} &
23 &  
\begin{block}
  \underline{bs \PASS \seqins~(bs)~(i)~(that)} \\ \PAND \\
  that.ao \PASS this
\end{block} &
\begin{block}
  \UPDATE \; t \; \begin{block}
                  \SET \; index = index +\mathrm{1} \\
                  \WHERE \; ao = this \; \AND \; index \geq i \texttt{;}
                  \end{block} \\
  \INSERT \; \INTO \; t \; (bs, ao, index) \; \VALUE \; (that, this, i) \textit{;}
\end{block} \\
\hline
\end{tabular}   
\vspace{-.2cm}
\caption{\small Assignment Transformation Pattern for \textit{sequence-to-optional} Bi-Association}\label{tab:pattern} 
\end{table*}}

From left to right, the columns of the table present declarations of
properties, numerical identifiers of patterns, their abstract
implementation in the substitution program, and their concrete
implementation in database queries.  The dummy variables $this$ and
$that$ are used to denote instances of, respectively, the current
class and the other end of the association. 

For each case (for each row of the completed table), we define a
transformation function {\small \verb|toSqlProc|} that turns a
substitution into a list of $\Sql$ query statements.
\begin{code}
  toSqlProc tm _ s@(Assign _ _) = transAssign tm s
  transAssign :: TABLE_MODEL -> Substitution -> [STATEMENT]
\end{code}
The function {\small \verb|toSqlProc|} delegates the task of
transforming base cases,~i.e. assignments, to another auxiliary
function \word{transAssign} that implements the $\mathrm{36}$
patterns. The recursive cases of \word{toSqlProc} are
straightforward. For example, to implement a guarded substitution, we
transform it into an {\small \verb|IfThenElse|} pattern that is
directly supported in the $\Sql$ domain; and to implement iterators
({\small \verb|ALL|}, {\small \verb|ANY|}), we instantiate a loop
pattern, declared with an explicit variant, that is guaranteed to
terminate.

\section{Correctness Proofs}

\label{sql:semantics}     
\label{sql:semantics:states} 
\label{sql:rel-struct:boo}
\label{sql:rel-struct:obj}

\noindent The correctness of both $\Booster$-to-$\Object$ and
$\Object$-to-$\Table$ transformations can be established by
constructing a relational model mapping identifiers and paths to
references and primitive values, and then showing that the different
reference mechanisms identify the same values in each case.  To prove
the correctness of the $\Table$-to-$\Sql$ transformation (shown as the
vertical, thick arrow in Figure~\ref{fig:sql:pipeline} on
page~\pageref{fig:sql:pipeline}), we need also to introduce
\textit{linking invariants} between model states.  We first formalise
states and operations for each model domain.  In
the Z notation, sets and relations may be described using a schema
notation, with separate declaration and constraint components and an
optional name: 
{\small\begin{schema}{name}
  declaration
  \where
  constraint
\end{schema}}\par\noindent
Either component may include schema references, with the special
reference $\Delta$ denoting two copies of another schema, typically
denoting before- and after-versions, the attributes of the latter
being decorated with a prime ($'$).  The remainder of the mathematical
notation is that of standard, typed, set theory.

We map the state $\Object$ model into a relational semantics
$\OBJState$, characterised by: 

{\small \begin{schema}{\OBJState}\label{revision:obj-state-schema}
  OBJECT\_MODEL \\
  extent : N\_CLASS \pfun \power ObjectId \\
  value  : ObjectId \pfun N\_PROPERTY \pfun Value
\where
  \dom extent = \dom class
  \\
  \forall c : N\_CLASS; o : ObjectId | \\ \t1 
    \block
    c \in \dom extent \land o \in extent~(c) @ 
      \dom~(value~(o)) = \dom~((class~c).property)
    \endblock
\end{schema}} 
\vspace*{-\baselineskip}
\par\noindent
The inclusion of $OBJECT\_MODEL$ (whose details are omitted here)
enables us to constrain the two mappings according to the type system
of the object model in question.  $Value$ denotes a structured type
that encompasses the possibilities of undefined value (for optional
properties), primitive value, and set and sequence of values.

The state of a table model will be composed of: 1) the type system of
the object model in context; and 2) functions for querying the state
of such a context object model.  More precisely, 
{\small \begin{schema}{TABLE\_MODEL}
  OBJECT\_MODEL \\
  nTableModel : N\_MODEL \\
  assocTables, setTables : \power (N\_CLASS \cross N\_PROPERTY) 
\end{schema}}    
\vspace*{-\baselineskip}
\par\noindent where $assocTables$, $setTables$ and $seqTables$ are
reflective queries: for example, $assocTables$ returns the set of
attributes/properties (and their context classes) that are stored in
the association tables.  We formalise the $\Table$ model state as:
{\small \begin{schema}{\TABState}
    \OBJState \\
    TABLE\_MODEL
\end{schema}}
\vspace*{-\baselineskip}
\par\noindent For each instance of $\OBJState$, there should be a
corresponding configuration of $TABLE\_MODEL$.  A $\Sql$ database
corresponds to a set of named tables, each containing a set of
column-value mappings:
\begin{trivlist}\item[]
  \begin{minipage}[b]{0.45\linewidth}   
    {\small \begin{schema}{\SQLState}
        tuples: N\_TABLE \pfun \power Tuple
      \end{schema}} 
  \end{minipage}
\begin{minipage}[b]{0.45\linewidth}
  {\small \begin{schema}{Tuple}
      values: N\_COLUMN \pfun ScalarValue
    \end{schema}}
\end{minipage}
\vspace*{-\baselineskip}
\end{trivlist}
We use $ScalarValue$ to denote the collection of basic types: strings,
integers, and Booleans.  We require mapping functions to retrieve
values from $\Table$ and $\Sql$:
\[
	\mappings : \TABState \cross (NClass \cross NProperty) \pfun \power (Value \cross Value)
\]
These return reference--value pairs for each kind of property.  For
example, set-valued properties are returned by
{\small 
  \[\setMappings == 
     \block
     \lambda s : \TABState ; p : NClass \cross NProperty @ \smallvskip \t1
         \bigcup \bigcbrace
                  o : ObjectId ; v : Value; vs : \power Primitive | \smallvskip \t1
                    o \in s.extent~(fst~p) \land  v = s.value~(o)~(snd~p) \land  v = setValue~(vs)
						@ \smallvskip \t2 
                    	\{~v' : vs @ \oidToSV{o} \mapsto \valToSV{v'}~\}
                        
                  \endbigcbrace
    \endblock\]}
\par\noindent
The set of mappings for a particular table is given by
{\small \[
    \block
        \lambda s : \SQLState ; n : NTable; c_1, c_2 : NColumn @ 
           \bigcbrace 
              row : s.tuples~(n) @ row.values~(c_1) \mapsto row.values~(c_2)
              \endbigcbrace
    \endblock
\]}
\vspace*{-\baselineskip}
\par\noindent 
and the necessary linking invariant is: {\small
\begin{schema}{\TabRelSql} 
	\TABState \\ 
	\SQLState
\where 
	C
\end{schema}
}   

\par\noindent where $C$ comprises six conjuncts, one for each
possible unordered combination of association end multiplicities.

Each operation is implemented as an atomic transaction.
$\OBJRelation$ represents the formal context, with the effect upon the
state being described as a binary relation ($\rel$).

{\small \begin{schema}{\OBJRelation}
  input : \power N\_VARIABLE \\
  output : \power N\_VARIABLE \\
  effect : (\OBJState \cross \OBJio) \rel (\OBJState \cross \OBJio)
  \where
  effect \in \OBJState \cross (input \fun Value) \rel \OBJState \cross
  (output \fun Value)
\end{schema}}
\vspace*{-\baselineskip}
\par\noindent Each element of $\OBJio == N\_VARIABLE \pfun Value$
represents a collection of inputs and outputs.  

Using $\OBJState$ and $\OBJRelation$, we may write $\OBJSystem$ to
denote the set of possible object system configurations, each
characterised through its current state (of type $\OBJState$) and its
set of indexed operations (of type $\OBJRelation$).  More precisely,
{\small \begin{schema}{\OBJSystem}
  state    : \OBJState \\
  relation : N\_CLASS \pfun N\_OPERATION \pfun \OBJRelation
\end{schema}} 
\vspace*{-\baselineskip}
\par\noindent We will describe the effect  of a
primitive assignment ($\PASS$), and use this as the basis for a
recursive definition of effect, based on the grammar of the GSL
notation.  If $AssignInput$ is the schema $[path? : \Path; e? :
Expression]$, then we may define
{\small\begin{schema}{AssignEffect}\label{revision:assigneffect}
    s, s' : \OBJState \\
    AssignInput \where
    s.nObjModel = s'.nObjModel \\
    s.sets = s'.sets \\
    s.classes = s'.classes
    \\
    s.extent = s'.extent
    \\
    \LET \block p == \target{path?}\; ; o == \context{path?} @ \block
    s'.value = \block
    s.value~\oplus \{ o \mapsto \\
    \t1 s.value~(o)~\oplus \{ p \mapsto eval~(e?)\}\}
           \endblock
         \endblock
       \endblock
\end{schema}}   
\vspace*{-\baselineskip}
\par\noindent
The input $path?$ can be either \word{OPATH} and \word{TPATH}: for the
former, the other input expression $e?$ involves paths, if any, of
type \word{OPATH}; for the latter, it is \word{TPATH}.  The ($\LET es
@ p$) expression, where $es$ consists of a list of
expression-to-variable bindings, denotes a predicate $p$ on the
variables of $es$.

We start by relating domains of the $\Object$ model and $\Table$
model, where assignment paths are specified in, respectively, {\small
  \verb|OPATH|} and {\small \verb|TPATH|}
(Fig~\ref{fig:sql:imp-proof}). In the $\Object$ model domain, an
assignment is parameterised by a path of type \word{BPATH} and an
expression that consists of paths, if any, of the consistent type. We
formalise each $\Object$ model assignment under the formal context of
$\OBJRelation$, by defining its $effect$ mapping though the constraint
of $AssignEffect$ and by requiring that the sets of external inputs
and outputs are empty.

{\small \begin{schema}{\AssignObj}
  \OBJRelation \\
  op? : \OPath \\
  oe? : Expression
\where
  \forall s, s' : \OBJState; AssignInput | 
    \block path? = OPath~(op?) \land e? = oe? @ 
      AssignEffect \iff (s, \{\}) \mapsto (s', \{\}) \in effect
    \endblock
\end{schema}}

The characterisation $\AssignTab$ of an assignment in the $\Table$
model domain is similar to that of $\AssignObj$, except that the
target is now of type \word{TPATH}, and the source is now of type
$Expression$.  We may then map our extended GSL substitution into a
relation:
\[
\GSLToRel{\_} : Substitution \fun ((\OBJState \cross \OBJio) \rel (\OBJState \cross \OBJio))
\] 
of the same type as the $effect$ component of $\OBJRelation$.  Given a
$\Table$ path $tp?$ and an expression $te?$, we represent the
assignment substitution $tp? \PASS te?$ by the effect relation of
$\AssignTab$ that exists uniquely with respect to $tp?$ and $te?$.
More precisely,
\[
\GSLToRel{tp? \PASS te?} = (\mu \AssignTab).effect
\]
where $\mu \AssignTab$ denotes the unique instance of $\TABState$ such
that the constraint as specified in $\AssignTab$ holds, and $.effect$
retrieves its relational effect on the model state. The definition of
$\AssignTab$ is very similar to that of $\AssignObj$, except that the
input path is constrained as $path? = TPath~(\dots)$.

We interpret a guarded substitution $g \PGUARD S$ as a relation that
has the same effect as $\smash{\GSLToRel{S}}$ within the domain of satisfying
states of guard $g$ (denoted as $\satisfyingOBJ{g}$); otherwise, it
just behaves like $\Pskip$ as it will be blocked and cannot achieve
anything. More precisely, we have:
\[
	\GSLToRel{g \PGUARD S} = \id~(\OBJState \cross \OBJio) \oplus (~\satisfyingOBJ{g} \dres \GSLToRel{S}~)
\]
Similar rules may be defined for other combinators.

Each transaction is composed of $\Sql$ queries, and similar to
$\OBJRelation$, we collect and produce, respectively, its list of
inputs and outputs upon its initiation and completion. We use
$\SQLRelation$ to denote such formal context, under which the
transformational effect on the state of database is defined
accordingly as a function, reflecting the fact that the database
implementation is deterministic in its effect.
{\small \begin{schema}{\SQLRelation}
  input, output : \power N\_VARIABLE \\
  effect : (\SQLState \cross \SQLio) \fun (\SQLState \cross \SQLio)
\where
  this \in input
\end{schema}}           
The mechanism of referencing the current object (via $this$) is
simulated through providing by default the value of $this$ for each
generated stored procedure or function. We model inputs and outputs in
the same way as we do for $\OBJio$, except that the range of values is
now of type $ScalarValue$.

For each $\Sql$ statement, we assign to it a relational semantics by
mapping it to a relation on states (of type $\SQLState$). This is a
similar process to that for $\GSLToRel{\_}$. More precisely, we
define:
\[
	\StmtToRel{\_} : Statement \fun ((\SQLState \cross \SQLio) \rel (\SQLState \cross \SQLio))
\]       
And since a $\Sql$ stored procedure is defined as a sequential composition, we also define
\[
	\StmtsToRel{\_} : \seq Statement \fun ((\SQLState \cross \SQLio) \rel (\SQLState \cross \SQLio))
\]    
to derive its effect through combining those of its component statements via relational composition.
For primitive query statements, we refer to their schema definitions. For example, we have:
\[
	\StmtToRel{\SQLUpdate{t}{sets}{cond}} = (\mu \UPDATE).effect
\]
where the state effect of query ($\SQLUpdate{table?}{sets?}{cond?}$)
is formally specified in a schema named $\UPDATE$. The $\UPDATE$ query
modifies in a table those tuples that satisfy a condition and takes as
inputs $table?$ a table name, $sets?$ a mapping that specifies how
relevant columns should be modified, and $cond?$ a Boolean condition
that chooses the range of tuples to be modified. The schema $\UPDATE$
is defined similarly as is $\AssignTab$, except that it imposes
constraints on the model state $\SQLState$.  We formalise
an $\SQLIf \dots \SQLThen \dots \SQLElse \dots$ statement as the union
of the semantic interpretations of the two sequences of statements in
its body, each suitably restricted on its domain. 

{\small
\[
  \StmtToRel{\SQLIfThenElse{b}{stmts_1}{stmts_2}} =
    \block
    (\satisfyingSQL{b} \dres \StmtsToRel{stmts_1})
    \; \cup \;
    (\satisfyingSQL{\NOT~b} \dres \StmtsToRel{stmts_2})
    \endblock
\]}

\noindent where $\satisfyingSQL{b}$ denotes the set of satisfying state of a $\Sql$ expression $b$. 

\smallskip

To define the semantics of a $\WHILE$ loop, we intend for the following equation to hold
{\small
\[
  \StmtToRel{\SQLWhile{b}{stmts}} = \StmtToRel{\SQLIfThenElse{b}{stmts \cat \langle \SQLWhile{b}{stmts} \rangle}{\langle \rangle}}             
\]}
\noindent where $\cat$ is the operator for sequence concatenation. By applying the definition of $\StmtToRel{\_}$ on \\
$\SQLIf \dots \SQLThen \dots \SQLElse \dots$ and $\StmtsToRel{\_}$ on $\langle \rangle$, we have
{\small
\[
  \fbox{$\StmtToRel{\SQLWhile{b}{stmts}}$}
  =  \block
     \satisfyingSQL{b} \dres (~\StmtsToRel{stmts} \comp \fbox{$\StmtToRel{\SQLWhile{b}{stmts}}$}~) \\
     \cup \\
     \satisfyingSQL{\NOT~b} \dres \id~(\SQLState \cross \SQLio)
     \endblock
\]}
Let us define a function
\[
  F~(X) = \block
          (~\satisfyingSQL{b} \dres (\StmtsToRel{stmts} \comp X)~) 
          \; \cup \;
          (~\satisfyingSQL{\NOT \; b} \dres \id~(\SQLState \cross \SQLio)~)
          \endblock
\]
When $X = \StmtToRel{\SQLWhile{b}{stmts}}$, we obtain
\[
  \StmtToRel{\SQLWhile{b}{stmts}} = F~(\; \StmtToRel{\SQLWhile{b}{stmts}} \;)
\]
which means that $\StmtToRel{\SQLWhile{b}{stmts}}$ should be a fixed-point of function $F$. The least fixed-point (LFP) of function $F$---i.e.~$\bigcup_{n \in \nat}F^{n}~(\emptyset)$---exists by Kleene's fixed-point theorem, since $F$ is easily provable to be continuous. We choose this LFP of $F$ for the value of $\StmtToRel{\SQLWhile{b}{stmts}}$.

\smallskip

\begin{figure}[htp]
  \centering %
  \includegraphics{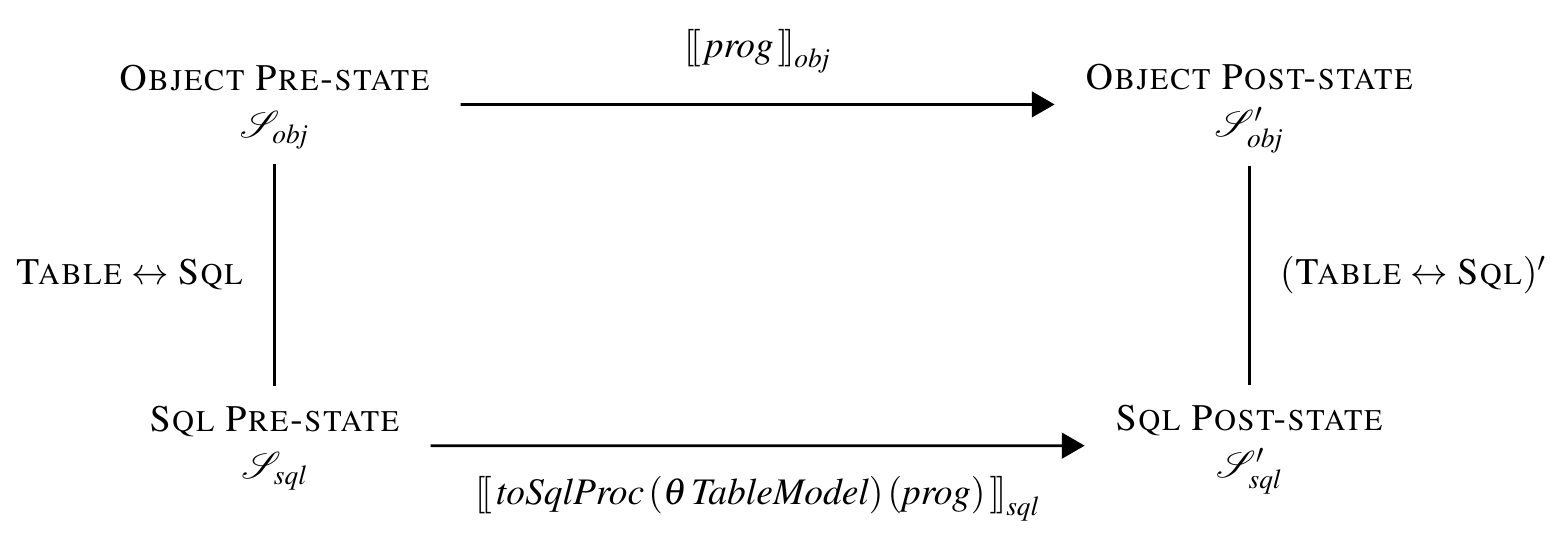}
  \vspace{-.2cm}
  \caption{\small Correctness of Model Transformation}
  \label{fig:sql:correctness} 
\end{figure}

We are now able to establish the correctness of the transformation with
respect to the linking invariant.  The commuting diagram of
Figure~\ref{fig:sql:correctness} shows how a substitution program
$prog$ and its context $\Table$ model (i.e.~$\theta TableModel$), are
mapped by the transformation $toSqlProc~(\theta~TableModel)~(prog)$ to
produce an $\Sql$ implementation.  The linking invariant holds for the
before states $\TabRelSql$ and for the after states $\TabRelSql'$.  We
then establish that for each state transformation, characterised by
the relational effect of the generated $\Sql$ code from $prog$, there
is at least a corresponding state transformation, characterised by the
relational effect of the $\Table$ program, $\GSLToRel{prog}$.  This is
an example of simulation between abstract data
types~\cite{DeRoever1999}.

We use a universal quantification ($\forall x | R(x) @ P(x)$) to state
our correctness criterion: the $x$ part declares variables, the $R(x)$
part constrains the range of state values, and the $P(x)$ part states
our concern. Schemas defined above (i.e.~$\OBJState$, $\SQLState$, and
$\TabRelSql$) are used as both declarations and predicates.   If we
declare
{\small
\begin{schema}{TransInput}
  TableModel \\
  prog: Substitution
\end{schema}
}
\vspace{-\baselineskip}
\par\noindent
to represent the inputs to the transformation, then
{\small \[
  \forall TransInput; \Delta \OBJState; \Delta \SQLState |
    \smallvskip \\
    \t1 \block
        \TabRelSql \land \StmtsToRel{toSqlProc~(\theta~TableModel)~(prog)} @ 
    \endblock   
    \bigbrace
      \exists \OBJState' @ \block
                            (\TabRelSql)' \land \GSLToRel{prog}
                           \endblock
      \endbigbrace
\]}         
\vspace{-\baselineskip}
\par\noindent     
With the relational semantics outlined above, we may establish this
result through a combination of case analysis and structural
induction.
       
\section{Example Implementation}
\label{sql:example}

\noindent Consider the implementation, on a relational database
platform, of the operation \word{reserve} introduced in
Section~\ref{background}.  Having translated the object model into a
collection of database tables, the generation process will produce a
stored procedure for each operation.  The guard for \word{reserve}
requires that the current number of allocations---characterised
through the cardinality of the set-valued attribute
\word{allocations}---is below a specific bound.  We might include such
a condition, for example, to ensure that the memory or storage
requirements of the system remain within particular bounds; this
may not be an issue for a hotel reservation system, but is a realistic
concern in critical systems development.  In the implementation, a
stored function is generated that will establish whether or not the
guard constraint holds for the current state, together with any input
values.  The remainder of the generated code will achieve the effect
specified in the original operation constraint, translated into the
representation, or orientation, of the database platform.

Class \word{Reservation} has \word{status} as an attribute, and this
is stored in the corresponding class table.  In the function,
\word{AUTO\_INCREMENT} allows the target $\Sql$ platform to generate a
unique identifier for each inserted row.  Set-valued properties, like
attribute \word{dates} in class \word{Reservation} are stored in
separate tables, with an \word{oid} column to identify the current
object in a given method call.  Associations such as \word{host} and
\word{reservations} are stored in separate tables, with an \word{oid}
column to identify the exact association instance.  Since attribute
\word{reservations} are also sequence-valued, an \word{index} column
is required. 
\begin{Verbatim}[frame=lines, label={Schema of Tables Updated by 'reserve'}, framesep=3mm, fontsize=\footnotesize, commandchars=\\\{\}, numbers=left, firstnumber=1, numbersep=3pt, numberblanklines=false, commentchar=!] 
CREATE TABLE `Reservation`(`oid` INTEGER AUTO_INCREMENT, PRIMARY KEY (`oid`), `status` CHAR(30)); 
CREATE TABLE `Room_reservations_Reservation_room`(`oid` INTEGER AUTO_INCREMENT, 
	PRIMARY KEY (`oid`), `reservations` INTEGER, `room` INTEGER, `index` INTEGER); 
\end{Verbatim}                          
We generate also integrity constraints for association tables:
although the generated procedures are guaranteed to preserve semantic
integrity, this affords some protection against errors in the design
of additional, manually-written procedures.

The value of the model-driven approach should be apparent following a
comparison of the original specification for \word{reserve} with the
fragments of the following $\Sql$ implementation.  Manual production
of queries that need to be take account of a large number of complex
model constraints---as well as, for examples, constraints arising from
data caching strategies---is time-consuming and error-prone.
Furthermore, we may expect the design of a system to evolve during
use: the challenge of maintaining correctness in the face of changing
specifications (and platforms) adds another dimension of complexity to
systems development; some degree of automation, in production and in
proof, is essential. 

In the following, variable names have been preserved from the
$\Booster$ domain, e.g.~the input and output parameters \word{dates?}
and \word{r!} at line 2, as well as caching variables
\word{`r!.status`}, \word{`r!.host`}, and \word{`r!.room`} at line 4.
Meta-variables are used to implement the \word{ALL} iterator in method
\word{reserve}: Line 5 declares, respectively, the bound variable
\word{`x`} and \word{`x\_variant`} the variant of the loop, and Line 6
declares a cursor over the set-valued input \word{dates?}.

\begin{Verbatim}[frame=lines, label={Queries Implementing 'reserve': Declarations}, framesep=3mm, fontsize=\footnotesize, commandchars=\\\{\}, numbers=left, firstnumber=1, numbersep=3pt, numberblanklines=false] 
CREATE PROCEDURE `Hotel_reserve`  (IN `this?` INTEGER, 
    IN `dates?` CHAR(30), IN `m?` INTEGER, OUT `r!` INTEGER)
BEGIN
  DECLARE `r!.status` CHAR(30); DECLARE `r!.host` INTEGER; DECLARE `r!.room` INTEGER;
  DECLARE `x` Date; DECLARE `x_variant` INTEGER;    
  DECLARE `x_cursor` CURSOR FOR (SELECT * FROM `dates?` WHERE TRUE);   
\end{Verbatim}  

Line 7 first creates a new instance of \word{Reservation} by
inserting, for output \word{r!}, a row formatted as $\langle oid,
\dots \rangle$ into the appropriate class table, where $oid$ is a
unique value generated by the built-in function
\word{last\_insert\_id()}, with the guarantee that each subsequent
call to this functions returns a new value. It then assigns this
unique identifier to \word{r!} for queries in later fragments to refer
to.
               
\begin{Verbatim}[frame=lines, label={Queries Implementing 'reserve': Creating an Empty Output}, framesep=3mm, fontsize=\footnotesize, commandchars=\\\{\}, numbers=left, firstnumber=7, numbersep=3pt, numberblanklines=false] 
    INSERT INTO `Reservation` () VALUE (); SET `r!` = last_insert_id ();                                                         
\end{Verbatim}

In Lines 8 to 10 the pair of \word{DROP TEMPORARY TABLE} and
\word{CREATE TEMPORARY TABLE} queries update the value of a cache
variable \word{`m?.reservations`} that denotes a multi-valued
property: this kind of caching is useful in large database
implementations.  In Line 11 we update the caching variable
\word{`r!.host`} of single-valued types of properties through a
\word{SELECT INTO} query.  We cache the value of attribute \word{host}
possessed by the reservation \word{r!}. Any later paths with
\word{`r!.host`} or \word{`m?.reservations`} as its prefix will be
able to use its value directly without re-evaluation.
   
\begin{Verbatim}[frame=lines, label={Queries Implementing 'reserve': Updating Caching Vars}, framesep=3mm, fontsize=\footnotesize, commandchars=\\\{\}, numbers=left, firstnumber=8, numbersep=3pt, numberblanklines=false]    
    DROP TEMPORARY TABLE IF EXISTS `m?.reservations`; 
    CREATE TEMPORARY TABLE `m?.reservations` AS 
      SELECT `reservations` FROM `Room_reservations_Reservation_room` WHERE `room` = `m?`;                      
    SELECT `status` INTO `r!.status` FROM `Reservation` WHERE `oid` = `r!`;  
\end{Verbatim} 

Lines 12 to 20 instantiate a finite loop pattern.  In Line 12 we
activate the declared cursor and and fetch its first available
value. In Line 13 we also calculate the size of the data set that the
cursor will iterate over and use it as the variant of the loop defined
in Lines 14 to 20. The exit condition (Line 14) is characterised
through decreasing---via the 2nd statement in Line 19---the value of
\word{x\_cursor}; the bound variable \word{x} is updated to the next
data item at the end of each iteration (via the 1st statement in Line
19). In each iteration of the loop, from Lines 15 to 17 we re-cache
the value of the set-valued path \word{r!.dates}, in case there are
other paths which contain it as a prefix and are used later in the
loop.  In Line 18 we perform the first substitution in the
specification of method \word{reserve}: we implement the substitution
{\small \verb| r!.dates := r!.dates \/ dates?|} via iterating through
the input \word{dates?} with a bound variable \word{`x`}.

\begin{Verbatim}[frame=lines, label={Queries Implementing 'reserve': Terminating Loop}, framesep=3mm, fontsize=\footnotesize, commandchars=\\\{\}, numbers=left, firstnumber=12, numbersep=3pt, numberblanklines=false] 
    OPEN `x_cursor`; FETCH `x_cursor` INTO `x`;   
    SELECT COUNT(*) INTO `x_variant` FROM `dates?` WHERE TRUE;   
    WHILE (`x_variant`) > (0) DO
      DROP TEMPORARY TABLE IF EXISTS `r!.dates`;
      CREATE TEMPORARY TABLE `r!.dates` AS 
        SELECT `dates` FROM `Reservation_dates` WHERE `oid` = `r!`;      
      INSERT INTO `Reservation_dates` (`oid`, `dates`) VALUE (`r!`, `x`);
      FETCH `x_cursor` INTO `x`; SET `x_variant` = `x_variant` - 1;
    END WHILE; CLOSE `x_cursor`;                                                     
\end{Verbatim}         

Line 21 implements the update \word{r!.status := unconfirmed}. The two
generated query statements---that are located in Lines 22 to 27 and
Lines 28 to 31---implement the last two parallel assignments in
\word{reserve} that update the \textit{optional-to-sequence}
association. They correspond exactly to the rules specified for
pattern 23 in Section~\ref{sql:trans:assignment}. The queries for the
middle two parallel assignments in \word{reserve}, updating the
\textit{one-to-sequence} association, are entirely similar. 

\begin{Verbatim}[frame=lines, label={Queries Implementing 'reserve': Performing Updates}, framesep=3mm, fontsize=\footnotesize, commandchars=\\\{\}, numbers=left, firstnumber=21, numbersep=3pt, numberblanklines=false]
  UPDATE `Reservation` SET `status` = 'unconfirmed' WHERE (`oid`) = (`r!`);
  UPDATE `Room_reservations_Reservation_room` 
  SET `index` = (`index`) + (1)
  WHERE `room` = `m?` AND 
        `index`>=(SELECT COUNT(`oid`)
                  FROM (SELECT `reservations` FROM `m?.reservations` WHERE TRUE) AS reservations
                  WHERE TRUE) + 1;     
  INSERT INTO `Room_reservations_Reservation_room` (`reservations`, `room`, `index`) VALUE
    (`r!`,`m?, (SELECT COUNT(`oid`)
                FROM (SELECT `reservations` FROM `m?.reservations` WHERE TRUE) AS reservations
                WHERE TRUE) + 1);    
\end{Verbatim}  

\section{Discussion} 

The principal contribution of this paper is the presentation of a
practical, formal, model-driven approach to the development of
critical systems.  Both the modelling notation and the target
programming language are given a formal, relational semantics: the
latter only for a specific subset of the language, sufficient for the
patterns of implementation produced by the code generation process.
The generation process is formalised as a functional program, easily
related to the corresponding transformation on the relational
semantics.  It is perfectly possible to prove the generator correct;
indeed, a degree of automatic proof could be applied here.  The task
of system verification is then reduced to the strictly simpler task of
model verification or validation.

The implementation platform chosen to demonstrate the approach is a
standard means of storing data, whether that data was originally
described in a hierarchical, a relational, or an object-oriented
schema.  In particular, there are many products that offer a means of
mapping~\cite{Russell2008} from object models (as used here) to a
relational database implementation: Hibernate~\cite{Hibernate} is
perhaps the best-known example.  However, translating the data
model to a data schema is relatively straightforward; the focus here is the
generation of correct implementations for operations.

At the same time, much of the work on program transformation is
focussed, unsurprisingly, upon code rewriting rather than the
generation of complete software components with persistent data.  The
work on Vu-X~\cite{Nakano2009}, where modifications to a web interface
are reflected back into the data model is an interesting exception,
but has yet to be extended to a formal treatment of data integrity.
The work on UnQL~\cite{Hidaka2009} supports the systematic development
of model transformation through the composition of graph-based
transformations: this is a powerful approach, but again no similar
framework has been proposed.

Some work has been done in precise data modelling in UML, for
example~\cite{Demuth1999}, but no formal account has been given for
the proposed translation of operations.  The Query/View/Transformation
approach~\cite{QVT2009} focuses on design models, but the
transformations~\cite{Jouault2008} are described in an imperative,
stateful, style, making proofs of correctness rather more difficult.
Recent work on generating provably correct code, for
example~\cite{Stenzel2011}, is restricted to producing primitive
getter and setter methods, as opposed to complex procedures.
Mammar~\cite{Mammar2008} adopts a formal approach to generating
relational databases from UML models.  However, this requires the
manual definition of appropriate guards for predefined update methods:
the automatic derivation of guards, and the automatic generation of
methods from arbitrary constraint specifications, as demonstrated
here, is not supported.  

The unified implementation and semantic framework for transformation
(Figure~\ref{fig:sql:pipeline}) presented here can be applied to any
modelling and programming notation that admits such a relational
semantics for the behaviour of components.  It is important to note
that the style of this semantics effectively limits the approach to
the development of sequential data components: that is, components in
which interactions with important data are managed as exclusive
transactions; our semantic treatment does not allow us to consider the
effects of two or more complex update operations executing
concurrently.  

In practice, this is not a significant limitation.  Where data is
encapsulated within a component, and is subject to complex business
rules and integrity constraints, we may expect to find locking or
caching protocols to enforce data consistency in the face of
concurrent requests, by means of an appropriate sequentialisation.
Where concurrency properties are important, they can be addressed
using process semantics and model-checking techniques; a degree of
automatic generation may even be possible, although this is likely to
be at the level of workflows, rather than data-intensive programs.

Work is continuing on the development of the transformation and
generation tools discussed here, with a particular emphasis upon the
incremental development of operation specifications and models.  It is
most often the case that a precise model will prove too restrictive:
when a property is written linking two or more attributes, it
constrains their interpretation; if one of these attributes is used
also elsewhere in the model, or within an operation, then that usage
may not always be consistent with the now formalised interpretation.
In our approach, such a problem manifests itself in the unavailability
of one or more operations, in particularly circumstances.

As a guard is generated for each operation, sufficient to protect any
data already acquired, each incremental version of the system can be
deployed without risk of data loss.  It can then be used in practice
and in earnest, allowing users to determine whether or not the
availability---or the overall design---of each operation and data view
matches their requirements and expectations.  Where an operation has a
non-trivial guard, additional analysis may be required to
demonstrate that the resulting availability matches requirements: in
many cases, the necessary check or test can be automated.  The work
described here provides a sound foundation for this development
process.   

\bibliographystyle{eptcs}
\bibliography{oxford-dphil}  

\end{document}

%% file: macros.tex
\def\Booster{\textsc{\mbox{Booster}}}
\def\Object{\textsc{\mbox{Object}}}
\def\Table{\textsc{\mbox{Table}}}
\def\Sql{\textsc{\mbox{Sql}}}
\def\PASS{\mathrel{\mathop:}=}
\def\PAND{\parallel}
\def\POR{\extchoice}
\def\PTHEN{\mathrel{;}}
\def\PGUARD{\longrightarrow}
\def\PEXISTS{\mbox{@}}
\def\PFORALL{\mbox{!}}

\def\Pass{\@ifnextchar~{\applyPass}{\_ \PASS \_}}
\def\applyPass~(#1, #2){#1 \PASS #2} 
\def\Pskip{skip}
\def\Pand{\@ifnextchar~{\applyPand}{\_ \PAND \_}}
\def\applyPand~(#1, #2){#1 \PAND #2}
\def\Por{\@ifnextchar~{\applyPor}{\_ \POR \_}}
\def\applyPor~(#1, #2){#1 \POR #2}
\def\Pthen{\@ifnextchar~{\applyPthen}{\_ \PTHEN \_}}
\def\applyPthen~(#1, #2){#1 \PTHEN #2}
\def\Pguard{\@ifnextchar~{\applyPguard}{\_ \PGUARD \_}}
\def\applyPguard~(#1, #2){#1 \PGUARD #2}
\def\Pexists{\@ifnextchar~{\applyPexists}{\PEXISTS\; \_\; :\; \_\; .\; \_}}
\def\applyPexists~(#1,#2,#3){\PEXISTS\; #1\; :\; #2\; .\; #3}
\def\Pforall{\@ifnextchar~{\applyPforall}{\PFORALL\; \_\; :\; \_\; .\; \_}}
\def\applyPforall~(#1,#2,#3){\PFORALL\; #1\; :\; #2\; .\; #3}   
\makeatother      
\def\Path{PATH}

\def\OPath{OPATH}

\def\SELECT{\texttt{SELECT}}

\def\UPDATE{\texttt{UPDATE}}
\def\INSERT{\texttt{INSERT}}
\def\INTO{\texttt{INTO}}
\def\VALUE{\texttt{VALUE}}

\def\SET{\texttt{SET}}
\def\WHERE{\texttt{WHERE}}

\def\SQLIf{\texttt{IF}}
\def\SQLThen{\texttt{THEN}}
\def\SQLElse{\texttt{ELSE}}
\def\WHILE{\texttt{WHILE}}
\def\DO{\texttt{DO}}
\def\END{\texttt{END}}

\def\NOT{\texttt{NOT}}
\def\AND{\texttt{AND}}

\def\OBJSystem{System_{obj}}

\def\State{{\cal{S}}}
\def\OBJState{\State_{obj}}
\def\TABState{\State_{tab}}
\def\SQLState{\State_{sql}}

\def\TabRelSql{\Table \rel \Sql}

\def\Relation{{\cal{R}}}
\def\OBJRelation{\Relation_{obj}}
\def\SQLRelation{\Relation_{sql}}
\newcommand{\decl}[1]{
  {\small \texttt{#1} }
}

\def\seqins~(#1)~(#2)~(#3){\mathrm{ins}~(#1, #2, #3)}
\def\seqdel~(#1)~(#2){\mathrm{del}~(#1, #2)}
\def\seqtake~(#1)~(#2){#1_{\dots #2}}
\def\seqdrop~(#1)~(#2){#1_{#2+\mathrm{1} \dots}}    

\def\mappings{{\cal{M}}}

\def\setMappings{\mappings_{set}}

\newcommand{\oidToSV}[1]{\lbag~#1~\rbag^{SV}} 

\newcommand{\valToSV}[1]{\lbag~#1~\rbag^{SV}}

\def\OBJio{IO_{obj}}
\def\SQLio{IO_{sql}} 

\def\TARGET{target}
\def\CONTEXT{context}
\newcommand{\target}[1]{\TARGET~\lbag~#1~\rbag}
\newcommand{\context}[1]{\CONTEXT~\lbag~#1~\rbag}

\def\TabRelSql{\Table \rel \Sql}

\def\Relation{{\cal{R}}}
\def\OBJRelation{\Relation_{obj}}
\def\SQLRelation{\Relation_{sql}}          

\def\AssignObj{Assign_{obj}}
\def\AssignTab{Assign_{tab}}

\newcommand{\GSLToRel}[1]{\lbag~#1~\rbag_{obj}}
\newcommand{\StmtToRel}[1]{\lbag~#1~\rbag_{sql}}
\newcommand{\StmtsToRel}[1]{\lbag~#1~\rbag_{\seq sql}}
\newcommand{\satisfyingOBJ}[1]{\lbag~#1~\rbag_{obj}^{states}}  
\newcommand{\satisfyingSQL}[1]{\lbag~#1~\rbag_{sql}^{states}}  

\newcommand{\SQLUpdate}[3]{\UPDATE \; #1 \; \SET \; #2 \; \WHERE \; #3}

\newcommand{\SQLIfThenElse}[3]{\SQLIf \; #1 \; \SQLThen \; #2 \; \SQLElse \; #3} 

\newcommand{\SQLWhile}[2]{\WHILE \; #1 \; \DO \; #2 \; \END \; \WHILE}

\usepackage{tabularx}
\usepackage{rotating}    
\usepackage{slashbox}
\usepackage[T1]{fontenc} 

\usepackage{relsize}  

\usepackage{fancyhdr} 
\usepackage{fancyvrb, courier} 
\renewcommand\ttdefault{txtt}

\newcommand{\word}[1]{{\small \texttt{#1}}}  
\DefineVerbatimEnvironment{code}{Verbatim}{frame=none, framesep=10mm, fontsize=\footnotesize}    
\DefineVerbatimEnvironment{booster}{Verbatim}{frame=none, framesep=10mm, fontsize=\footnotesize, commandchars=~\[\]}

\def\smallvskip{\smallskip \\}

\def\bigbrace{\left(
        \begin{array}{c}
        \begin{block}}
\def\endbigbrace{\end{block}
        \end{array} 
    \right)}
          
\def\bigbrace{\left(
        \begin{array}{c}
        \begin{block}}
\def\endbigbrace{\end{block}
        \end{array} 
    \right)}

\def\bigcbrace{\left\{
		\begin{array}{c}
	    \begin{block}}
\def\endbigcbrace{\end{block}
	    \end{array} 
	\right\}}